# A Census of Exoplanets in Orbits Beyond 0.5 AU via Space-based Microlensing

White Paper for the Astro2010 PSF Science Frontier Panel

David P. Bennett[1]

(phone: 574-631-8298, email: bennett@nd.edu),

J. Anderson[2], J.-P. Beaulieu[3], I. Bond[4], E. Cheng[5], K. Cook[6], S. Friedman[2], B.S. Gaudi[7], A. Gould[7], J. Jenkins[8], R. Kimble[9], D. Lin[10], J. Mather[9], M. Rich[11], K. Sahu[2], T. Sumi[12], D. Tenerelli[13], A. Udalski[14], and P. Yock[15]

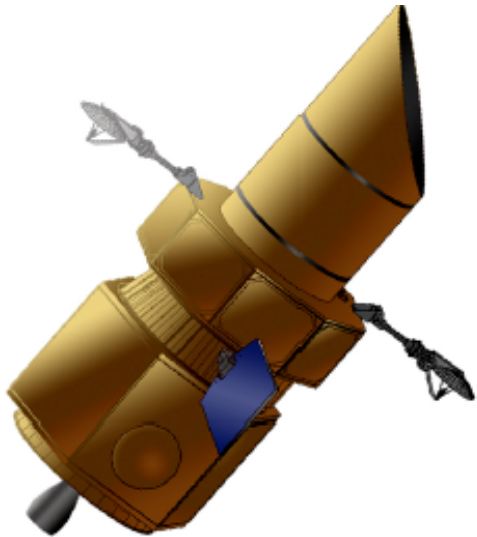 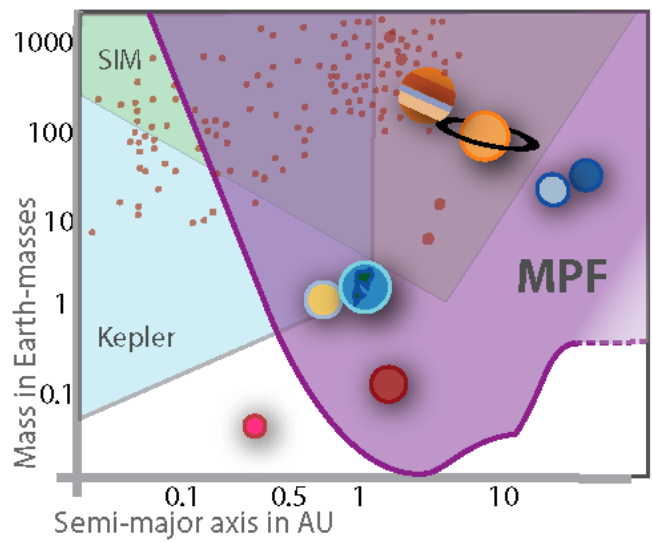

[1] University of Notre Dame, Notre Dame, IN, USA
[2] Space Telescope Science Institute, Baltimore, MD, USA
[3] Institut d'Astrophysique, Paris, France
[4] Massey University, Auckland, New Zealand
[5] Conceptual Analytics, LLC, Glen Dale, MD, USA
[6] Lawrence Livermore National Laboratory, USA
[7] Ohio State University, Columbus, OH, USA
[8] SETI Institute, Mountain View, CA, USA
[9] NASA/Goddard Space Flight Center, Greenbelt, MD, USA
[10] University of California, Santa Cruz, CA, USA
[11] University of California, Los Angeles, CA, USA
[12] Nagoya University, Nagoya, Japan
[13] Lockheed Martin Space Systems Co., Sunnyvale, CA, USA
[14] Warsaw University, Warsaw, Poland
[15] University of Auckland, Auckland, New Zealand

## ABSTRACT

A space-based gravitational microlensing exoplanet survey will provide a statistical census of exoplanets with masses $\geq 0.1 M_\oplus$ and orbital separations ranging from 0.5AU to $\infty$. This includes analogs to all the Solar System's planets except for Mercury, as well as most types of planets predicted by planet formation theories. Such a survey will provide results on the frequency of planets around all types of stars except those with short lifetimes. Close-in planets with separations < 0.5 AU are invisible to a space-based microlensing survey, but these can be found by Kepler. Other methods, including ground-based microlensing, cannot approach the comprehensive statistics on the mass and semi-major axis distribution of extrasolar planets that a space-based microlensing survey will provide. The terrestrial planet sensitivity of a ground-based microlensing survey is limited to the vicinity of the Einstein radius at 2-3 AU, and space-based imaging is needed to identify and determine the mass of the planetary host stars for the vast majority of planets discovered by microlensing. Thus, a space-based microlensing survey is likely to be the only way to gain a comprehensive understanding of the architecture of planetary systems, which is needed to understand planet formation and habitability. The proposed Microlensing Planet Finder (MPF) mission is an example of a space-based microlensing survey that can accomplish these objectives with proven technology and a cost of under $300 million (excluding launch vehicle).

## 1. Basics of the Gravitational Microlensing Method

The physical basis of microlensing is the gravitational attraction of light rays by a star or planet. As illustrated in Fig. 1, if a "lens star" passes close to the line of sight to a more distant source star, the gravitational field of the lens star will deflect the light rays from the source star. The gravitational bending effect of the lens star "splits", distorts, and magnifies the images of the source star. For Galactic microlensing, the image separation is $\leq 4$ mas, so the observer sees a microlensing event as a transient brightening of the source as the lens star's proper motion moves it across the line of sight.

Gravitational microlensing events are characterized by the Einstein ring radius,

$$R_E = 2.0 \text{ AU} \sqrt{\frac{M_L}{0.5 M_\odot} \frac{D_L (D_S - D_L)}{D_S (1 \text{ kpc})}} ,$$

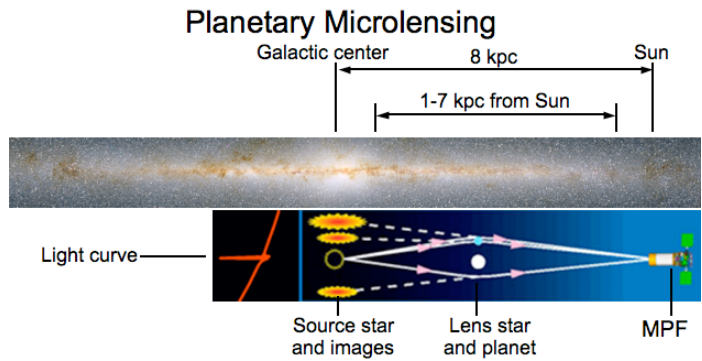

**Fig. 1:** *The geometry of a microlensing planet search towards the Galactic bulge. Main sequence stars in the bulge are monitored for magnification due to gravitational lensing by foreground stars and planets in the Galactic disk and bulge.*

where $M_L$ is the lens star mass, and $D_L$ and $D_S$ are the distances to the lens and source, respectively. This is the radius of the ring image that is seen with perfect alignment between the lens and source stars. The lensing magnification is determined by the alignment of the lens and source stars measured in units of $R_E$, so even low-mass lenses can give rise to high magnification microlensing events. The duration of a microlensing event is given by the Einstein ring crossing time, which is typically 1-3 months for stellar lenses and a few days or less for a planet.

**Planets are detected via light curve deviations** that differ from the normal stellar lens light curves (Mao & Paczynski 1991). Usually, the signal occurs when one of the two images due to lensing by the host star passes close to the location of the planet, as indicated in Fig. 1 (Gould & Loeb 1992), but planets can also be detected at very high magnification where the gravitational field of the planet destroys the symmetry of the Einstein ring (Griest & Safizadeh 1998).

## 2. Capabilities of the Microlensing Method

**Planets down to one tenth of an Earth mass can be detected**. The probability of a detectable planetary signal and its duration both scale as $R_E \sim M_p^{1/2}$, but given the optimum alignment, planetary signals from low-mass planets can be quite strong. The limiting mass for the microlensing method occurs when the planetary Einstein radius becomes smaller than the projected radius of the source star (Bennett & Rhie 1996). The ~5.5 $M_\oplus$ planet detected by Beaulieu et al. (2006) is near this limit for a giant source star, but most microlensing events have G or K-dwarf source stars with radii that are at least 10 times smaller than this. So, the sensitivity of the microlensing method extends down to $< 0.1 M_\oplus$, as the results of a detailed simulation of the MPF mission (Bennett & Rhie 2002) show in Fig. 2.

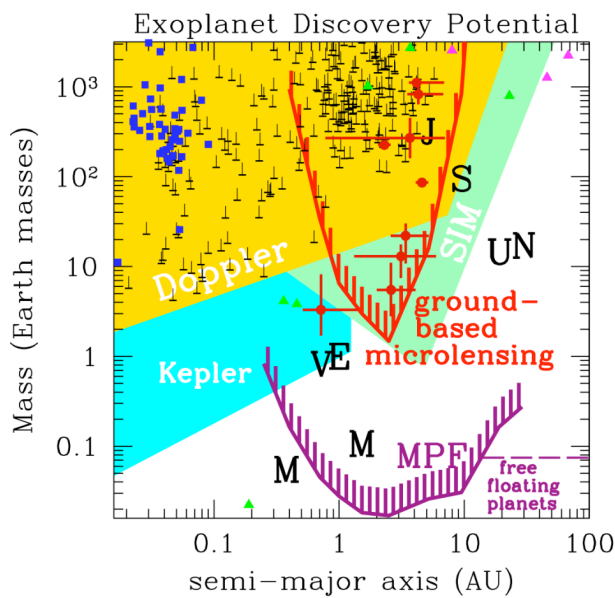

*Fig. 2: Space-based microlensing (MPF) is sensitive to planets above the purple curve in the mass vs. semi-major axis plane. The gold, green and cyan regions indicate the sensitivities of radial velocity surveys, SIM and Kepler, respectively. Our Solar System's planets are indicated by their fist initials, and the known extrasolar planets are shown. Ground-based microlensing discoveries are in red; Doppler detections are inverted T's; transit detections are blue squares; timing and imaging detections are green and magenta triangles, respectively.*

**Microlensing is sensitive to a wide range of planet-star separations and host star types.** The host stars for planets detected by microlensing are a random sample of stars that happen to pass close to the line-of-sight to the source stars in the Galactic bulge, so all common types of stars are surveyed, including G, K, and M-dwarfs, as well as white dwarfs and brown dwarfs. Microlensing is most sensitive to planets at a separation of $\sim R_E$ (usually 2-3 AU) due to the strong stellar lens magnification at this separation, but the sensitivity extends to arbitrarily large separations. It is only planets well inside $R_E$ that are missed because the stellar lens images that would be distorted by these inner planets have very low magnifications and a very small contribution to the total brightness. These features can be seen in Fig. 2, which compares the sensitivity of a space microlensing mission (MPF) with expectations for other planned and current programs. Other ongoing and planned programs can detect, at most, analogs of two of the Solar System's planets, while a space-based microlensing survey can detect seven—all but Mercury. The only method with comparable sensitivity to MPF is the Kepler space-based transit survey, which complements the microlensing method with sensitivity at semi-major axes, $a \le 1$

AU. The sensitivities of MPF and Kepler overlap at separations of ~1 AU, which corresponds to the habitable zone for G and K stars.

The red crosses in Fig. 2 indicate the 5 gas giant (Bond et al. 2004; Udalski et al. 2005; Gaudi et al. 2008; Dong et al. 2008) and four ~$10 M_\oplus$ "super-earth" planets discovered by ground-based microlensing (Beaulieu et al. 2006; Gould et al. 2006; Bennett et al. 2008; Sumi et al. in preparation 2009). A preliminary analysis suggests that about one third of all stars are likely to have a super-earth at 1.5-4AU whereas radial velocity surveys find that only about 3% of stars have gas giants in this region (Butler et al. 2006).

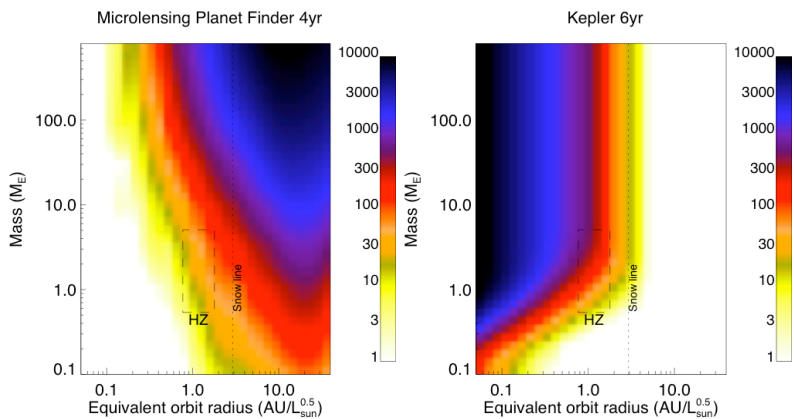

**Fig. 3:** Microlensing Planet Finder (MPF) vs Kepler exoplanet "depth of survey" estimates from the Exoplanet Task Force report (Lunine et al. 2008). The shading indicates the number of stars that are effectively searched for exoplanets as a function of mass and orbital radius, scaled by stellar luminosity so that the HZs of all types of stars are at ~ 1 AU.

Fig. 3 shows a comparison of the depth or survey estimates made by the Exoplanet Task Force, and this shows that a 4-year space microlensing mission finds a similar number of habitable zone (HZ) planets as an extended Kepler mission. Almost all the HZ planets found by microlensing orbit G and K stars.

**Microlensing light curves yield unambiguous planet parameters.** For the great majority of events, the basic planet parameters (planet:star mass ratio, planet-star separation) can be "read off" the planetary deviation (Gould & Loeb 1992; Bennett & Rhie 1996; Wambsganss 1997). Possible ambiguities in the interpretation of planetary microlensing events have been studied in detail (Gaudi & Gould 1997; Gaudi 1998), and these can be resolved with high quality, continuous light curves that will be routinely acquired with a space-based microlensing survey. A space-based survey will also detect most of the planetary host stars, which generally allows the host star mass, approximate spectral type, and the planetary mass and separation to be determined (Bennett et al. 2007; Bennett 2009). The distance to the planetary system is determined when the host star is identified, so a space-based microlensing survey will also measure how the properties of exoplanet systems change as a function of distance from the Galactic Center. There is usually some redundancy in the measurements that determine the properties of the host stars, and so the determination is robust to complicating factors, such as a binary companion to the background source star.

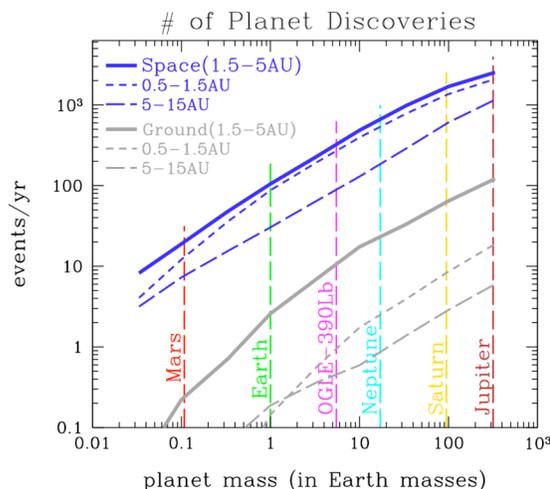

**Fig. 4:** The expected number of MPF planet discoveries as a function of the planet mass if every star has a single planet in the given separation of ranges.

**Detailed simulations indicate a large number of planet detections.** Bennett & Rhie (2002) and Gaudi (unpublished) have independently simulated space-based microlensing surveys. These simulations included variations in the assumed mission capabilities that allow us to explore how changes in the mission design will affect the scientific output, and they form the basis of our predictions in Figs. 2-5. Fig. 4 shows the expected number of planets that MPF would detect at orbital separations of 0.5-1.5, 1.5-5, and 5-15 AU. This plot assumes an average of one planet per star in each range of separations. Only a space-based survey has a significant detection rate for Earths in all 3 separation ranges, and so only a space-based survey can sample a wide range of separations including the HZ (see Fig. 3). Fig. 5 shows the expected detection rate for free-floating planets assuming one such planet per star. Free-floating planets are expected to be a common by-product of most planet formation scenarios (Levison et al. 1998; Goldreich et al. 2004), and only a space-based microlensing survey can detect free-floating planets of $\leq 1M_\oplus$, which are important evidence of a crucial step in the planet formation process.

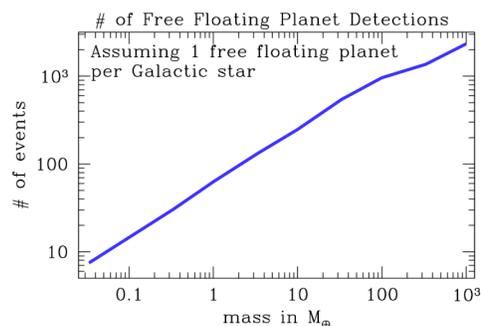

**Fig. 5:** *The expected number of MPF free-floating planet discoveries.*

## 3. A Space-based Microlensing Survey Is Needed

Microlensing relies upon the high density of source and lens stars towards the Galactic bulge to generate the stellar alignments that are needed to generate microlensing events, but this high star density also means that the bulge main sequence source stars are not generally resolved in ground-based images, as Fig. 6 demonstrates. This means that the precise photometry needed to detect planets of $\leq 1M_\oplus$ is not possible from the ground unless the magnification due to the stellar lens is moderately high. This, in turn, implies that ground-based microlensing is only sensitive to terrestrial planets located close to the Einstein ring (at ~2-3 AU). The full sensitivity to terrestrial planets in all orbits from 0.5 AU to ∞ comes only from a space-based survey.

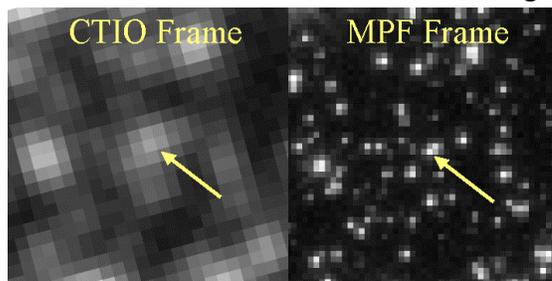

**Fig. 6:** *A comparison between an image of the same star field in the Galactic bulge from CTIO in 1" seeing and a simulated MPF frame (based on an HST image). The indicated star is a microlensed main sequence source star.*

**Planetary host star detection from space yields precise star and planet parameters.** For all but a small fraction of planetary microlensing events, space-based imaging is needed to detect the planetary host stars, and the detection of the host stars allows the star and planet masses and separation in physical units to be determined. This can be accomplished with HST observations for a small number of planetary microlensing events (Bennett et al. 2006), but space-based survey data will be needed for the detection of host stars for the thousands of planetary microlensing events that we expect from a space mission. Fig. 7 shows the distribution of planetary host star masses and the predicted uncertainties in the masses and separation of the planets and their host stars (Bennett et al. 2007) from simulations of the MPF mission. The host stars with masses determined to better than 20% are indicated by the red histogram in Fig. 7(a), and these are primarily the host stars that can be detected in MPF images.

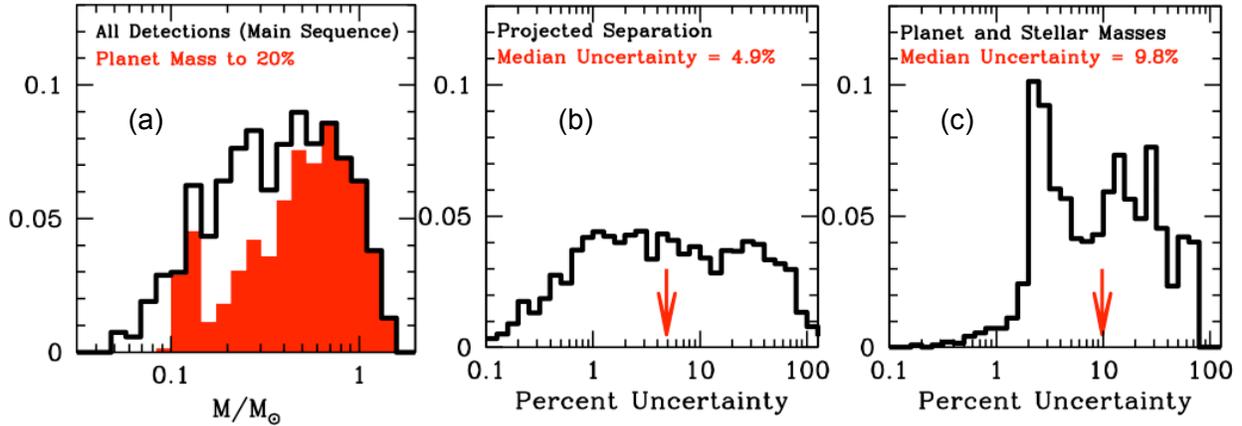

*Fig. 7:* (a) The simulated distribution of stellar masses for stars with detected terrestrial planets. The red histogram indicates the subset of this distribution for which the masses can be determined to better than 20%. (b) The distribution of uncertainties in the projected star-planet separation. (c) The distribution of uncertainties in the star and planet masses.

Ground-based microlensing surveys also suffer significant losses in data coverage and quality due to poor weather and seeing. As a result, a significant fraction of the planetary deviations seen in a ground-based microlensing survey will have poorly constrained planet parameters due to poor light curve coverage (Peale 2003).

## 4. A Space-Based Microlensing Survey Constrains Planet Formation Theories

Rapid advancement in exoplanet research is driven by both extensive observational searches around mature stars as well as the construction of planet formation and evolution models. Perhaps the most surprising discovery so far is the great diversity in the planets' dynamical properties, but these results are largely confined to planets that are unusually massive or reside in very close orbits. The core accretion theory suggests most planets are much less massive than gas giants and that the critical region for understanding planet formation is the "snow-line", located in the region (1.5-4 AU) of greatest microlensing sensitivity (Ida & Lin 2005; Kennedy et al. 2006). Early results from ground-based microlensing searches (Beaulieu et al. 2006; Gould et al. 2006) appear to confirm these expectations. A space-based microlensing survey would extend the current sensitivity of the microlensing method down to masses of ~$0.1M_\oplus$ over a large range (0.5AU-∞) in separation, and in combination with Kepler, such a mission provides sensitivity to sub-Earth mass planets at all separations. The semi-major axis region probed by space-microlensing provides a cleaner test of planet formation theories than the close-in planets detected by other methods, because planets discovered at > 0.5 AU are more likely to have formed *in situ* than the close-in planets. The sensitivity region for space-microlensing includes the outer habitable zone for G and K stars through the "snow-line" and beyond, and the lower sensitivity limit reaches the regime of planetary embryos at ~$0.1M_\oplus$. It may be that such planets are much more common than planets of $1M_\oplus$ because their type-1 migration time is much longer.

**Space-microlensing tests core accretion.** The space-microlensing census of low-mass planets should also provide direct evidence of features of the proto-planetary disk predicted by the core accretion theory. There are several physical processes that control the development of planetary embryos and planets in the proto-planetary disk. In the inner disk, the size of planetary embryos is controlled by the isolation mass, and the isolation mass is expected to jump by an order of magnitude across the "snow-line" because of the increased surface density of solids in the disk. But the number of gas giant and super-earth planets is also expected to increase beyond the snow line, while the planetary growth time increases. This means that it is more likely for the growth of outer planets to be terminated via gravitational scattering of planetesimals or the protoplanets themselves. Scattering would also result in the removal of lower mass planets into very distant orbits or even out of the gravitational influence of the host stars altogether, but space-based microlensing can still detect planets in these locations. The frequency of planets of different masses and separations that a space-based microlensing survey provides will yield a unique insight into the planetary formation process and will allow us to determine the importance of these processes.

**The habitability of a planet depends on its formation history.** The suitability of a planet for life depends on a number of factors, such as the average surface temperature, which determines if the planet resides in the habitable zone. However, there are many other factors that also may be important, such as the presence of sufficient water and other volatile compounds necessary for life (Raymond et al. 2004; Lissauer 2007). Thus, a reasonable understanding of planet formation is an important foundation for the search for nearby habitable planets and life.

## 5. Implementation of a Space-based Microlensing Mission

A space-based microlensing mission requires a space telescope of at least 1m-aperture, with a focal plane of > 0.5 sq. deg. in the near IR (or visible) with an orbit with a continuous view the Galactic bulge. It requires no new technology, and can be accomplished with a budget of less than $300 million (excluding the launch vehicle). The Microlensing Planet Finder or MPF (shown on the cover page) is an example of such a mission (Bennett et al. 2004), which has been proposed to NASA's Discovery program. Another, very similar, design known as DUNE (for Dark Universe Explorer) had been proposed to CNES and ESA to study dark energy via the weak lensing method (Refregier et al. 2008). This remarkable similarity between these designs suggests that a joint mission could be even more cost effective.

## 6. Endorsement by the Exoplanet Task Force

The Exoplanet Task Force (ExoPTF) recently released a report (Lunine et al. 2008) that evaluated all of the current and proposed methods to find and study exoplanets, and they expressed strong support for space-based microlensing. Their finding regarding space-based microlensing states that: *"Space-based microlensing is the optimal approach to providing a true statistical census of planetary systems in the Galaxy, over a range of likely semi-major axes, and can likely be conducted with a Discovery-class mission."* Their conclusion that a space-based microlensing survey can be conducted with a Discovery-class mission (i.e. a cost ≤ $300 million excluding the launch vehicle), is in agreement with the judgment of the 2006 Discovery review panel, which found that the proposed Microlensing Planet Finder costs were credible. The ExoPTF's support of space-based microlensing is also reflected in their recommendation B.II.2, which states: *"Without impacting the launch schedule of the astrometric mission cited above, launch a Discovery-class space-based microlensing mission to determine the statistics of planetary mass and the separation of planets from their host stars as a function of stellar type and location in the galaxy, and to derive $\eta_\oplus$ over a very large sample."*

## 7. Discussion

A space-based microlensing survey provides a census of extrasolar planets that is complete (in a statistical sense) down to $0.1 M_\oplus$ at orbital separations ≥ 0.5 AU, and when combined with the results of the Kepler mission a space-based microlensing survey will give a comprehensive picture of all types of extrasolar planets with masses down to well below an Earth mass. This complete coverage of planets at all separations can be used to calibrate the poorly understood theory of planetary migration. This fundamental exoplanet census data is needed to gain a comprehensive understanding of processes of planet formation and migration, and this understanding of planet formation is an important ingredient for the understanding of the requirements for habitable planets and the development of life on extrasolar planets.

A subset of the science goals can be accomplished with an enhanced ground-based microlensing program (Gaudi et al. 2009), which would be sensitive to Earth-mass planets in the vicinity of the "snow-line". But such a survey would have its sensitivity to Earth-like planets limited to a narrow range of semi-major axes, so it would not provide the complete picture of the frequency of exoplanets down to $0.1 M_\oplus$ that a space-based microlensing survey would provide. Furthermore, a ground-based survey would not be able to detect the planetary host stars for most of the events, and so it will not provide the systematic data on the variation of exoplanet properties as a function of host star type that a space-based survey will provide.

## 8. Conclusions

The past 15 years have seen the emergence of the study of extrasolar planets as a new sub-field of astronomy, but while many planets have been discovered, we are still ignorant about many of the basic characteristics of extrasolar planet systems. We are only beginning to answer the most basic question about planetary systems: What is the basic architecture of extrasolar planetary systems and the stars that host them? There are many projects that can begin to address this question, but only a space-based microlensing mission can complete the census of exoplanets beyond 0.5 AU. Our knowledge of exoplanets and their formation will remain seriously incomplete until such a mission is flown.